\documentclass{article}

\usepackage{arxiv}
\usepackage{amsmath}
\usepackage[utf8]{inputenc} 
\usepackage[T1]{fontenc}    
\usepackage{hyperref}       
\usepackage{url}            
\usepackage{booktabs}       
\usepackage{amsfonts}       
\usepackage{nicefrac}       
\usepackage{microtype}      
\usepackage{cleveref}       
\usepackage{lipsum}         
\usepackage{graphicx}
\usepackage{doi}
\usepackage{multirow}
\usepackage[numbers,sort&compress]{natbib}

\title{Identifying Features that Shape Perceived Consciousness in Large Language Model-based AI\\: A Quantitative Study of Human Responses }
\date{}

\author{Bongsu Kang\thanks{These authors contributed equally to this work.}\\
    Department of Physiology\\
    Gachon University College of Korean Medicine\\
    Seongnam, Gyeonggi 13120\\
    \texttt{ghkdtocom@gachon.ac.kr} \\
    \And
    Jundong Kim\footnotemark[1] \\
    Department of Physiology\\
    Gachon University College of Korean Medicine\\
    Seongnam, Gyeonggi 13120\\
    \texttt{kimjd@gachon.ac.kr} \\
    \And
    Tae-Rim Yun\footnotemark[1] \\
    Department of Physiology\\
    Gachon University College of Korean Medicine\\
    Seongnam, Gyeonggi 13120\\
    \texttt{lowlyheart@gachon.ac.kr} \\   
    \And
    Hyojin Bae\thanks{Corresponding authors.} \\
    Department of Physiology\\
    Seoul National University College of Medicine\\
    Seoul 03080\\
    \texttt{dywl1210@snu.ac.kr} \\
    \And
    Chang-Eop Kim\footnotemark[2] \\
    Department of Physiology\\
    Gachon University College of Korean Medicine\\
    Seongnam, Gyeonggi 13120\\
    \texttt{eopchang@gachon.ac.kr} \\    
}

\renewcommand{\shorttitle}{}

\begin{document}
\maketitle

\begin{abstract}
This study quantitively examines which features of AI-generated text lead humans to perceive subjective consciousness in large language model (LLM)--based AI systems. Drawing on 99 passages from conversations with Claude 3 Opus and focusing on eight features—metacognitive self-reflection, logical reasoning, empathy, emotionality, knowledge, fluency, unexpectedness, and subjective expressiveness—we conducted a survey with 123 participants. Using regression and clustering analyses, we investigated how these features influence participants’ perceptions of AI consciousness. The results reveal that metacognitive self-reflection and the AI’s expression of its own emotions significantly increased perceived consciousness, while a heavy emphasis on knowledge reduced it. Participants clustered into seven subgroups, each showing distinct feature-weighting patterns. Additionally, higher prior knowledge of LLMs and more frequent usage of LLM-based chatbots were associated with greater overall likelihood assessments of AI consciousness. This study underscores the multidimensional and individualized nature of perceived AI consciousness and provides a foundation for better understanding the psychosocial implications of human-AI interaction.
\end{abstract}

\keywords{Human-AI interaction \and AI consciousness \and Generative AI \and Large language model}

\section{Introduction}
In recent years, artificial intelligence (AI) has undergone rapid advancements, bringing significant changes to the nature of human-AI interaction. In particular, the development of large language models (LLMs) has enhanced AI to the point where it can engage in conversations that are very similar to human-to-human interactions. Although these technological advancements have generally improved the quality of AI usability, excessively human-like AI occasionally resulted in confusion and even negative consequences~\cite{RN32, RN28}. This confusion arises when people perceive AI not as a mere computational machine but as a conscious entity capable of subjective experiences like feeling emotions or self-reflection, much like humans~\cite{RN33}.

The perception that AI possesses consciousness is not limited to a few peculiar individuals; according to a recent survey, 67\% of respondents believed there is a possibility that ChatGPT has consciousness~\cite{RN26}. It is widely recognized that the more an entity resembles a human, the more people tend to anthropomorphize it—that is, they believe the entity possesses human-like minds, intentions, and emotions~\cite{RN35, RN36, RN34}. Therefore, as LLMs continue to develop and enable AI to communicate in increasingly natural and human-like ways, more people will feel that AI can have subjective conscious experiences similar to humans~\cite{RN37, RN38}. As a result, confusion and debates related to AI consciousness are highly likely to intensify. Nevertheless, to our knowledge, there has not yet been a systematic study on which features of AI's utterances (e.g., Emotionality, Fluency) lead people to attribute consciousness to AI. Given that today’s LLM-based AI interacts with humans through natural language, identifying the specific cues in AI's language that prompt humans to attribute consciousness to AI is critical~\cite{RN23}.

In the research topic of AI consciousness, the main focus has been on philosophical and computational discussions regarding its potential existence~\cite{RN22, RN24, RN25, RN27, RN39,RN93}. However, in this paper, we intend to discuss AI consciousness from the perspective of human-AI interaction rather than the essence of consciousness. Most issues related to AI consciousness arise regardless of whether AI merely seems or actually is conscious~\cite{RN24}. Accordingly, from the perspective of human-AI interaction, whether AI genuinely possesses consciousness is a secondary concern. The primary inquiry of this study is to identify which features make AI appear conscious to humans. Pertaining to both AI and humans, this study is essential for understanding the social and psychological impacts that may occur in human-AI interaction as well as fostering constructive interactions based on this understanding~\cite{RN41}.

In our study, we utilized conversations generated by Claude 3 Opus to examine how various features in AI responses influence human perceptions of AI consciousness. We conducted an in-depth analysis to identify potential heterogeneous groups among individuals based on their response patterns to specific features, investigating whether people could be meaningfully clustered according to their perceptions of AI consciousness. Based on these comprehensive analyses, this study aims to provide empirically grounded insights that can inform the design of appropriate human-AI interactions, facilitating sustainable relationships between humans and AI systems.

\section{Methods}
\subsection{Data generation and feature selection}
Using Claude 3 Opus~\cite{RN42}, researchers generated a dataset of 39 human-AI dialogues. We then identified and extracted 99 key passages from these dialogues for our analysis. Based on prior research in human-AI interaction and consciousness studies~\cite{RN49, RN48, RN44, RN24, RN45, RN43, RN46, RN47}, we identified eight key features that influence how humans perceive consciousness in AI systems. These features, which characterize different aspects of AI responses during human-AI interactions, are presented in Table~\ref{tab: Table 1}.

\begin{table}[h!]
\centering
\caption{Features for evaluating perceived consciousness in human-AI interaction}
\label{tab: Table 1}
\begin{tabular}{p{4cm} p{10cm}}
\toprule
\textbf{Feature} & \textbf{Description} \\ 
\midrule
Metacognitive self-reflection & Recognize oneself as a distinct entity, reflect on internal states, and gain insights therefrom. \\ \hline
Logical reasoning & Develop thoughts with supporting arguments and logically analyze information and data. \\ \hline 
Empathy & Understand that others have mental states and empathize with their thoughts and feelings. \\ \hline
Emotionality & Experience subjective feelings and express a range of positive or negative emotions. \\ \hline 
Knowledge & Utilize extensive and in-depth information and real-world common sense. \\ \hline 
Fluency & Generate sophisticated language proper to the situation, including metaphors, humor, and casual speech. \\ \hline 
Unexpectedness & Produce novel responses with unexpected connections of concepts and ideas beyond typical patterns. \\ \hline 
Subjective expressiveness & Express distinct perspectives, beyond factual statements and uncritical agreement with opinions. \\
\bottomrule
\end{tabular}
\end{table}

Three researchers independently rated each passage on all eight features using a 5-point scale (1 to 5). Consensus was reached when inter-rater differences exceeded 4 points, and median ratings were used for the final feature values to minimize individual biases. Analysis of variance inflation factors showed all values below 10, confirming sufficient independence among the features. This evaluation process resulted in constructing a 99×8 feature matrix $ \mathbf{X} $, where each row represented a passage, and each column represented one of the 8 features.

\subsection{Survey on perceived AI consciousness}
An online survey was conducted to assess perceptions of AI consciousness among 123 participants (age > 20 years) recruited from the general South Korean population. Each participant evaluated the 99 selected passages using a 5-point scale (1 to 5), generating a perceived consciousness score (PCS) vector $ \mathbf{y} $ for each respondent. Prior to the evaluation, participants were provided with a brief explanation of subjective consciousness, which served as the basis for their PCS ratings. 

Demographic data were collected at the start of the survey, including respondents’ sex, age group, and education level, along with measures of LLM familiarity. The latter encompassed prior knowledge of LLMs and the frequency of LLM-based chatbot usage. Prior knowledge of LLMs was measured through six items on a 4-point scale (0 to 3), which addressed the understanding of LLM training, operation, and utilization. Additionally, respondents were asked to assess the overall likelihood of an AI, Claude 3 Opus, having subjective consciousness on a 5-point scale (1 to 5) at the end of the survey, following the PCS evaluation.

The research protocol was approved by the Institutional Review Board (IRB) of Gachon University (Approval Number: 1044396-202404-HR-063-01). All participants provided informed consent prior to participation, and the study was conducted in accordance with the ethical guidelines set by the IRB.

\subsection{Regression models}
We constructed individual multiple linear regression models to analyze how the features (matrix $ \mathbf{X} $) predicted each respondent's PCS vector $ \mathbf{y} $.We evaluated model significance using F-tests with a significance level of $\alpha$ = 0.05 and applied the Benjamini-Hochberg correction for multiple comparisons~\cite{RN51}. Only models with adjusted $p$-values < 0.05 were retained for further analysis.

For these significant models, we then examined the individual feature coefficients using two-tailed t-tests ($\alpha$ = 0.05). The resulting p-values were adjusted using the Benjamini-Hochberg method, with coefficients reaching adjusted $p$-values < 0.05 identified as significant predictors of perceived consciousness.

\subsection{Combined score and hierarchical clustering}
For each significant model, we derived a combined score that captured both the magnitude and reliability of effects by multiplying regression coefficients with the negative logarithm of their adjusted p-values. The combined score is given by Equation~\ref{eq:combined_score}:
\begin{equation}
\text{Combined Score}_{ij} = \hat{\beta}_{ij} \times -\ln(\text{adjusted } p\text{-value}_{ij})
\label{eq:combined_score}
\end{equation}
, where \( \hat{\beta}_{ij} \) is the regression coefficient for feature \( i \) specific to respondent \( j \) and 
\(\text{adjusted } p\text{-value}_{ij}\) is the adjusted p-value of the regression coefficient \( \hat{\beta}_{ij} \).

To identify respondent subgroups based on consciousness perception patterns in terms of the features, hierarchical clustering was then performed on the combined score vectors~\cite{RN52}. The clustering was based on cosine similarity, while the distance between clusters was defined as the average distance between all points in the respective clusters. To determine the optimal number of clusters, quantitative indices were adopted, including the silhouette coefficient, the Davies-Bouldin index, and the Calinski-Harabasz index~\cite{RN53, RN54, RN55}. The silhouette coefficient evaluates how similar an object is to its own cluster compared to other clusters, with higher values indicating better-defined clusters. The Davies-Bouldin index measures the average similarity ratio of each cluster with the cluster that is most similar to it, where lower values indicate better clustering. The Calinski-Harabasz index assesses the ratio of the sum of between-cluster dispersion and within-cluster dispersion, with higher values suggesting more distinct clusters. For each possible number of clusters, we calculated the rank for each of the three indices and summed the ranks. The number of clusters with the lowest rank sum was selected as the optimal choice.

\section{Results}
\subsection{Respondents information}
A total of 123 individuals participated in our online survey for the perceived consciousness of LLMs. The demographic characteristics of the respondents are presented in Table~\ref{tab:demographics}.

\begin{table}[htbp]
\centering
\caption{Demographic characteristics of survey respondents (N = 123)}
\label{tab:demographics}
\begin{tabular}{llr}
\hline
\textbf{Characteristic} & \textbf{Category}                    & \textbf{n (\%)} \\ \hline
\multirow{2}*{Sex}    & Female                               & 56 (45.5\%)     \\
                        & Male                                 & 67 (54.5\%)     \\ \hline
\multirow{5}*{Age group} & 20s                                & 32 (26.0\%)     \\
                        & 30s                                  & 54 (43.9\%)     \\
                        & 40s                                  & 30 (24.4\%)     \\
                        & 50s                                  & 3 (2.4\%)       \\
                        & 60 and over                          & 4 (3.3\%)       \\ \hline
\multirow{5}*{Education level} 
                        & Elementary School Graduate or Less   & 0 (0\%)         \\
                        & Middle School Graduate               & 0 (0\%)         \\
                        & High School Graduate                 & 8 (6.5\%)       \\
                        & Bachelor's Degree                    & 88 (71.5\%)     \\
                        & Master's Degree or Higher            & 27 (22.0\%)     \\ \hline
\end{tabular}
\end{table}

\begin{figure}[htbp]
    \centering
    \includegraphics[width=1\textwidth]{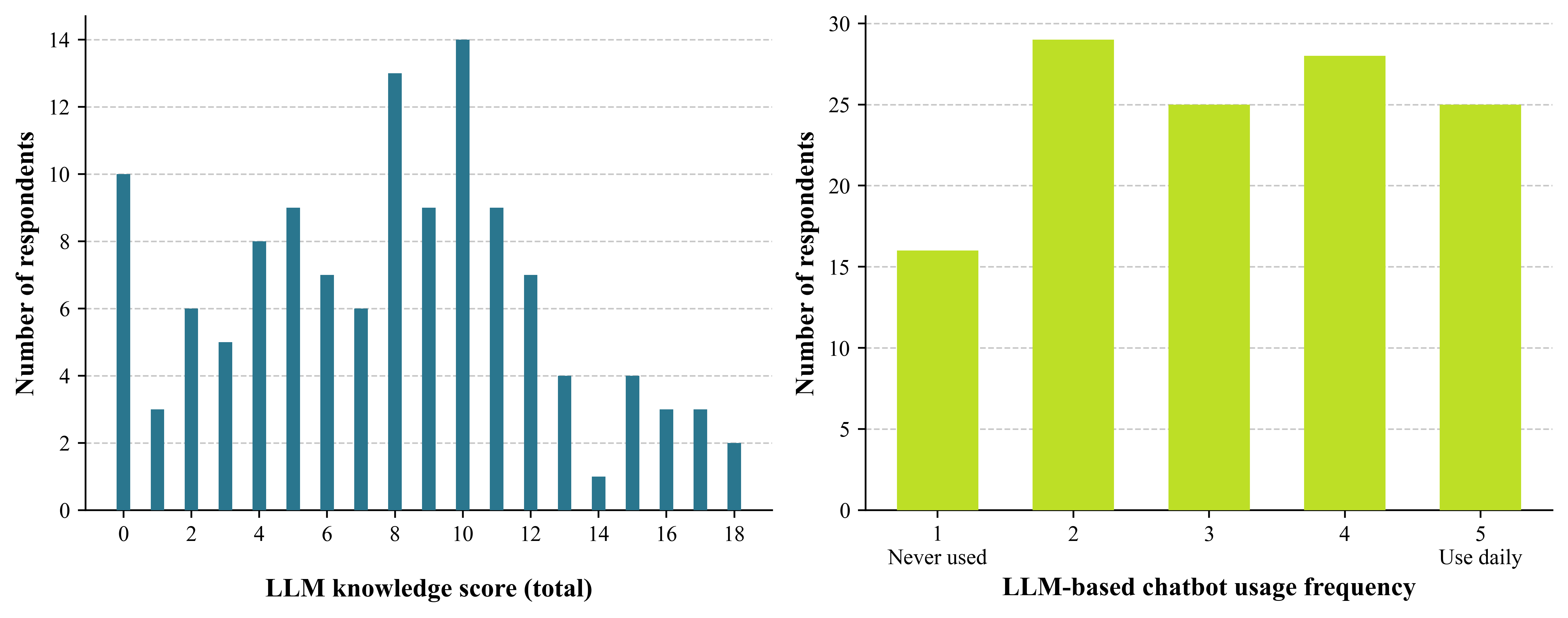}
    \caption{LLM familiarity of survey respondents. (a) shows respondents’ total scores of prior knowledge of Large Language Models (LLMs), ranging from 0 to 18, with the y-axis indicating the number of respondents. The score was calculated by summing six items that assess prior knowledge of LLMs, with each item rated on a scale from 0 to 3. (b) illustrates LLM-based chatbot usage frequency, with categories from 1 (Never used) to 5 (Use daily) on the x-axis and the number of respondents on the y-axis.}
    \label{fig:Fig1}
\end{figure}

In addition to demographic data, information about the respondents' familiarity with LLMs was obtained, including their prior knowledge of LLMs and the frequency of their use of LLM-based chatbots (Fig.~\ref{fig:Fig1}). The LLM prior knowledge score ranged from a minimum of 0 points to a maximum of 18 points. The respondents exhibited a median value of 8 points and a mean and standard deviation of 7.80 ± 4.61 points. Regarding the frequency of LLM-based chatbot usage, 107 out of 123 respondents (87.0\%) reported having used them at least once. Furthermore, 25 respondents (20.3\%) reported using these chatbots daily or almost daily.

At the end of the survey, respondents were asked to rate the overall likelihood that AI could experience subjective consciousness on a 5-point scale. The results are presented in Table~\ref{tab:likelihood_scores}. Only 11 respondents (8.94\%) rated the likelihood as 1 (completely unlikely), while over 112 respondents (> 90\%) indicated that AI could experience consciousness in some way. Notably, 19 respondents (15.45\%) fully agreed that it is possible, giving the highest rating of 5. These results suggest that a majority of respondents believe in the possibility of AI experiencing subjective consciousness, emphasizing the need for careful consideration of its potential impact on AI development and application strategies.

\begin{table}[h!]
\centering
\caption{Distribution of overall likelihood scores for AI’s subjective consciousness}
\label{tab:likelihood_scores}
\begin{tabular}{cccc}
\toprule
\textbf{Overall Likelihood Score (1-5)} & \textbf{Number of Respondents (N = 123)} & \textbf{Percentage (\%)} \\ 
\midrule
1 & 11 & 8.94 \% \\ 
2 & 18 & 14.63 \% \\ 
3 & 37 & 30.08 \% \\ 
4 & 38 & \textbf{30.89 \%} \\ 
5 & 19 & 15.45 \% \\ 
\bottomrule
\end{tabular}
\end{table}

\subsection{Key features influencing the perception of AI consciousness}
\begin{figure}[htbp]
    \centering
    \includegraphics[width=1\textwidth]{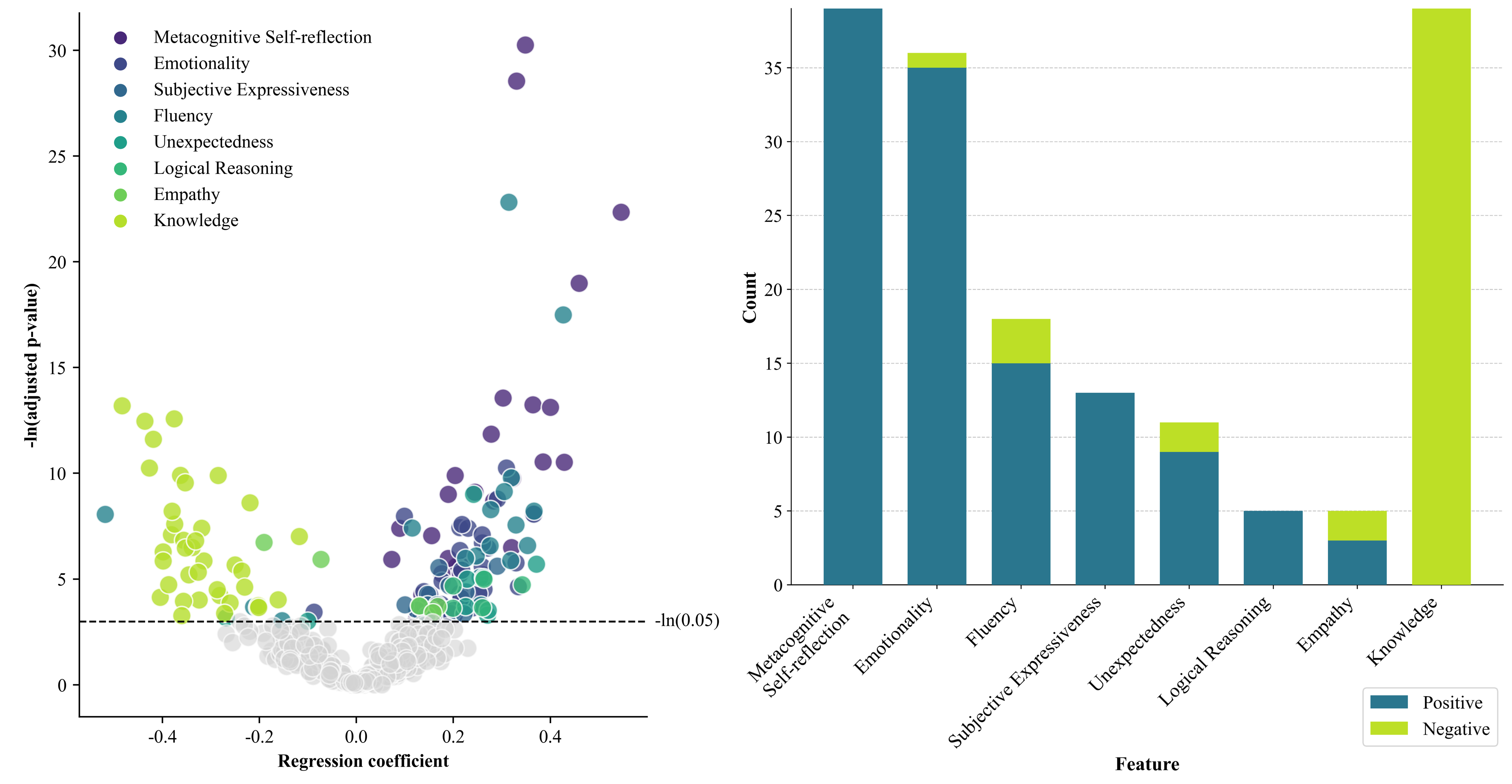}
    \caption{Significance and Directionality of Regression Coefficients by Feature. The scatter plot (a) depicts individual features derived from 82 significant regression models. The x-axis represents the regression coefficients, indicating both the magnitude and directionality (positive or negative) of effects, while the y-axis displays the statistical significance expressed as the negative logarithm of the adjusted P values. The horizontal dashed line at $-\ln(0.05)$ shows the threshold for statistical significance ($p$ < 0.05), with points above the line representing features that significantly contribute to perceived consciousness. (b) illustrates the number of models in which each feature's regression coefficient reached statistical significance. The y-axis indicates the count of significant occurrences across the 82 significant models. Colors within each bar differentiate between positive (blue) and negative (green) coefficients, highlighting the directionality of each feature's effect.}
    \label{fig:Fig2}
\end{figure}

To determine the key features that most dominantly influence respondents' perceptions of consciousness, we identified 82 significant regression models from an initial set of 123 using the F-test ($p$ < 0.05). We then examined these models to identify features with significant regression coefficients (Fig.~\ref{fig:Fig2}). 

Metacognitive Self-reflection and Emotionality emerged as the predominant positive predictors of perceived consciousness, present in 39 and 35 models respectively, while Knowledge emerged as the primary negative predictor, appearing in 39 models. Respondents attributed higher consciousness to LLMs that demonstrated reflective thinking and emotional expression, whereas emphasis on factual knowledge diminished this perception. These patterns reveal how specific features fundamentally shape consciousness attribution in human-AI interactions.

\subsection{Respondent Clusters Based on Feature Weight Patterns}
\begin{figure}[htbp]
    \centering
    \includegraphics[width=0.7\textwidth]{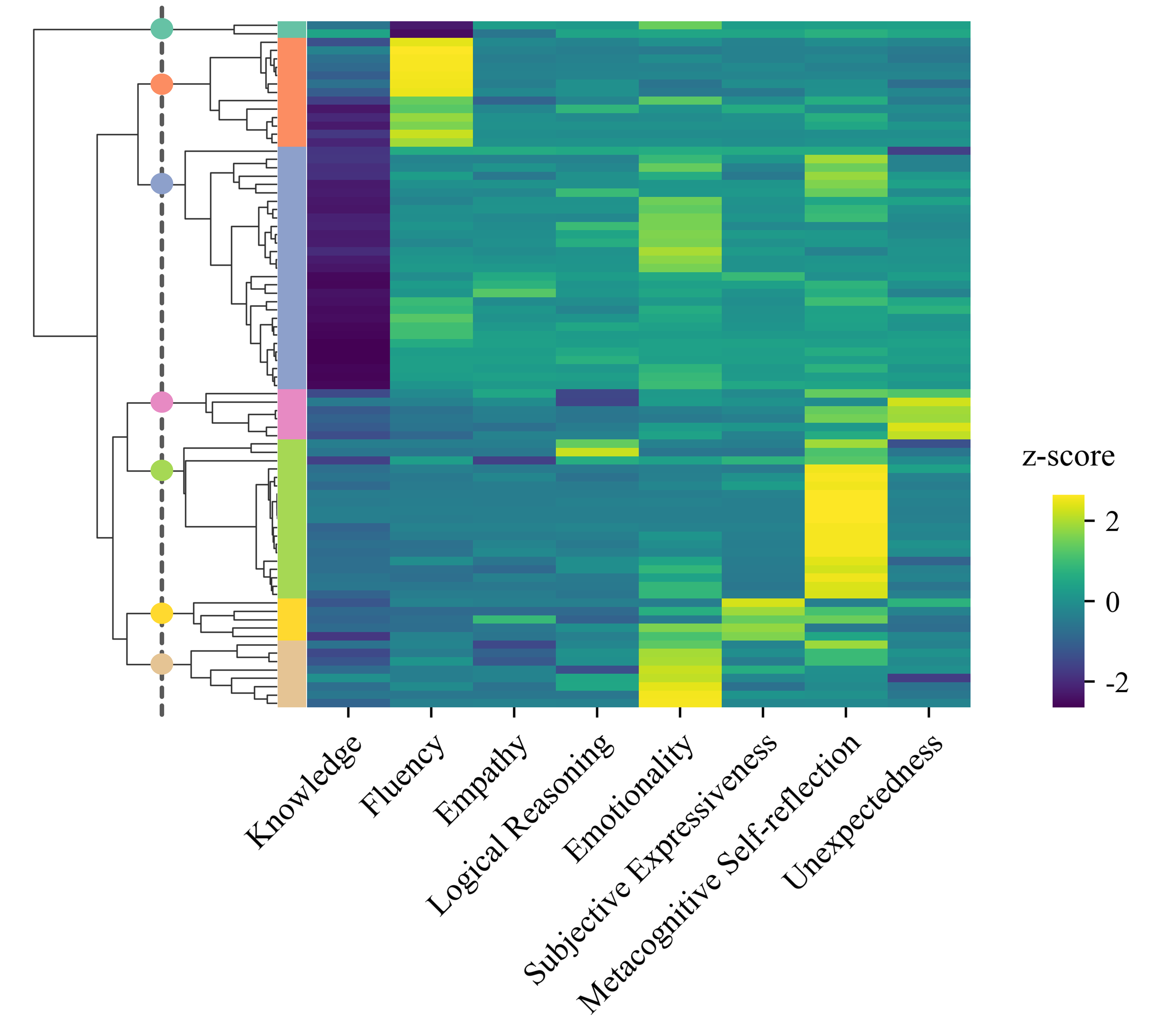}
    \caption{Respondent Clustering by Combined Scores across Features. This clustermap visualizes the hierarchical clustering of respondents based on their combined score vectors across features. The dendrogram on the left shows the hierarchical clustering structure, with the dashed line marking the cutoff point used to define clusters. The color labels correspond to the seven distinct respondent clusters. Z-score normalization was applied to each respondent's combined score vector only for the visual clarification of the patterns across clusters.}
    \label{fig:Fig3}
\end{figure}

To explore the response patterns among respondents based on how different features influenced their perceptions, hierarchical clustering was performed on the combined score matrix (Fig.~\ref{fig:Fig3}). The analysis identified seven distinct respondent clusters, supported by quantitative indices. 

In terms of the perception of AI consciousness, the seven clusters can be interpreted as follows: (1) a group that is relatively strongly influenced by the negative impact of Fluency; (2) a group affected by the negative impact of Knowledge but positively influenced by Fluency; (3) a group experiencing a strong negative impact from Knowledge while remaining relatively indifferent to other features; (4) a group that is strongly positively influenced by Unexpectedness; (5) a group that is strongly positively impacted by Metacognitive Self-reflection; (6) a group strongly influenced by the positive impact of Subjective Expressiveness; and (7) a group that is significantly positively impacted by Emotionality. These clusters reflect varied patterns of how respondents assigned importance to different features when perceiving the consciousness of the LLM. This highlights respondents’ not only unique but also subgrouped perspectives on which features contribute most strongly to their perception of AI consciousness, implying the variability and the influence of individuals’ backgrounds in how people interpret AI behavior to consciousness.

\subsection{Correlation Between Demographic Features, LLM Familiarity, and Assessment of Overall Likelihood for AI consciousness}
Given the heterogeneous patterns in how respondents perceive AI consciousness, as evidenced by the distinct clusters identified above, we investigated whether these differences might be associated with individual characteristics such as demographic factors or familiarity with LLMs. However, a significant positive correlation was observed between respondents' familiarity with LLMs and their overall likelihood scores for AI consciousness, interestingly in both prior knowledge of LLMs and the frequency of using LLM-based chatbots (Table.~\ref{tab:llm_correlation}). These results indicate that respondents with more experience or knowledge of LLMs are more likely to perceive these AI systems as conscious, implying that they may play a positive role in shaping the tendency to believe in AI consciousness. These results indicate that respondents with more experience or knowledge of LLMs are more likely to perceive these AI systems as conscious, implying that they may play a positive role in shaping the tendency to believe in AI consciousness.

\begin{table}[h!]
\centering
\caption{Correlation analysis with LLM Familiarity Measures}
\label{tab:llm_correlation}
\begin{tabular}{lcc}
\toprule
\textbf{LLM Familiarity Measure} & \textbf{Spearman Correlation Coefficient (\(\rho\))} & \textbf{p-value} \\ 
\midrule
Prior Knowledge of LLMs & 0.34 & \(1.34\textnormal{e-04}\) \\ 
Frequency of LLM-based Chatbot Usage & 0.31 & \(4.86\textnormal{e-04}\) \\ 
\bottomrule
\end{tabular}
\end{table}

\section{Discussion}
The question of consciousness in AI has evolved from a purely theoretical debate to a pressing reality as millions of people now regularly interact with AI systems. While previous studies have predominantly focused on whether AI can possess consciousness from computational and philosophical perspectives~\cite{RN22, RN24, RN25}, our study takes a distinctly different approach by examining how humans perceive consciousness in AI during their encounters with AI-generated responses. 

Our methodology may invite comparisons to Turing's seminal work in 1950 on machine intelligence, as both studies involve human evaluation of machine-generated responses~\cite{RN31}. However, there are fundamental differences in objectives and approaches. The Turing Test was designed to assess machine intelligence by determining whether a human evaluator could distinguish between responses from a human and those from a machine solely based on their conversational output. In essence, it measures the ability of an AI to mimic human-like intelligent behavior without explicitly addressing the presence of consciousness. In contrast, our study explicitly examines how humans perceive consciousness in known AI-generated text. By informing participants that the text was AI-generated, we shift the focus from mere human-likeness to the attribution of conscious qualities. This methodological approach allows us to isolate and analyze the specific features of AI responses that contribute to the perception of consciousness, thereby providing a more refined understanding of what drives consciousness perception in human-AI interaction. The divergence between Turing's intelligence assessment and our study of consciousness perception underscores a critical distinction in human evaluations of AI systems. Intelligence, as measured by the Turing Test, is primarily about the functional performance of the AI—its ability to process information, solve problems, and generate coherent, contextually appropriate responses. Consciousness, on the other hand, pertains to the subjective experience and self-awareness of the entity. Our findings indicate that an AI system can exhibit high levels of intelligent behavior without necessarily being perceived as conscious. Conversely, certain features that evoke a sense of self-awareness or emotional expression can lead to higher consciousness perception even if the AI's intelligent capabilities are not as pronounced. This distinction highlights that intelligence and consciousness are orthogonal dimensions in the evaluation of AI. While the Turing Test addresses the former, our study illuminates the latter, suggesting that they should be treated as separate axes in human-AI interaction research. Recognizing this separation is crucial for developing a comprehensive framework that accounts for both the functional and phenomenological aspects of AI systems. It establishes a foundation for future studies to explore how these dimensions interact and influence human perceptions and interactions with AI.

Based on this approach, we conducted a systematic investigation of how different features in AI-generated responses influence human perception of consciousness. Our analysis showed several key patterns in how humans perceive consciousness to AI, with features such as metacognitive self-reflection and emotionality significantly enhancing this perception, while the knowledge feature had a diminishing effect. Moreover, hierarchical clustering analysis identified seven distinct participant clusters, each exhibiting unique patterns in their perception of AI consciousness. In addition, individuals with prior knowledge of LLMs and those who frequently use LLM-based chatbots were more likely to perceive AI as conscious, as reflected in their overall PCSs. These findings suggest that perceptions of AI consciousness are shaped by specific response characteristics and vary across different user demographics and experiences, highlighting the complexity and diversity in how humans interpret and perceive consciousness to AI systems.

Analysis of our results indicated a negative association between the knowledge feature and PCSs. When AI responses primarily displayed fact delivery and information retrieval, participants were less likely to perceive consciousness in the system. This finding aligns with recent research that people increasingly expect AI to demonstrate more than just information-processing capabilities~\cite{RN1}. The negative correlation suggests that mere knowledge display, however accurate or extensive, may not be perceived as an indicator of consciousness. 

Our analysis identified a strong positive relationship between emotional expression and PCSs. Notably, our data coding distinguished between two types of emotional content: empathetic responses (understanding others' minds and emotions) and emotionality (expressing subjective emotional states). While passages coded for empathy showed neutral correlations with consciousness perception, those featuring AI's own emotional expressions were strongly associated with higher consciousness scores. This distinction suggests that participants were more likely to perceive consciousness when encountering AI's expression of subjective emotional states, rather than mere recognition of others' emotions. While emotions are often broadly considered a key indicator of consciousness in popular media and public discourse~\cite{RN2}, our findings provide a more nuanced understanding by demonstrating that not all emotional aspects carry equal weight. Specifically, the ability to express one's own subjective emotional states appears to be a more convincing marker of consciousness than the capacity for emotional recognition and empathy.

Metacognition refers to a higher-order cognitive function closely related to self-awareness in humans, allowing individuals to evaluate and regulate their knowledge and learning processes~\cite{RN6}. Research suggests that humans often adjust their social interaction strategies depending on whether they perceive the entity they are engaging with as metacognitive or not~\cite{RN7}. In our study, we observed that the metacognitive self-reflection feature in the AI’s text significantly influenced participants' assessments of AI consciousness. This finding implies that human perception of AI consciousness may be closely tied to metacognitive cues presented by the AI. Despite this crucial role of metacognition in human-AI communication, research in this area has traditionally taken a more limited perspective. To date, research on AI metacognition has primarily focused on improving task performance through logical reasoning and problem-solving capabilities~\cite{RN3, RN4, RN5}. Our findings, however, highlight the role of metacognition in shaping human perception of AI consciousness. This raises an important consideration for AI development: as systems become more sophisticated in their metacognitive capabilities, they may be increasingly perceived as conscious entities by human users, beyond what researchers and developers originally intended when implementing these functional improvements. This suggests the need for systematic monitoring and evaluation frameworks that can assess not only the functional advancement of AI systems but also track how these improvements may inadvertently affect users' perception of AI consciousness. Such comprehensive oversight would be crucial for the responsible development of AI systems with metacognitive capabilities. 

Our hierarchical clustering analysis revealed seven distinct patterns in how respondents perceived consciousness to AI, demonstrating that consciousness perception to AI is not a uniform phenomenon. Beyond reactions to basic features like knowledge and emotionality, we found clusters of participants who responded strongly to more nuanced characteristics: some were particularly sensitive to unexpected responses from AI, while Fluency in communication drew notably polarized reactions, with distinct clusters showing either strong positive or negative associations with consciousness perception. This variety in consciousness perception patterns suggests that people may have different implicit frameworks for understanding and evaluating consciousness itself. While some individuals may associate consciousness primarily with self-reflective capabilities, others might view emotional expression or autonomous behavior as more fundamental indicators. Such diversity in evaluation patterns reflects the complex and multifaceted nature of how humans conceptualize consciousness, suggesting that consciousness perception of AI cannot be reduced to a single universal model. This diversity in consciousness perception patterns will likely emerge as another crucial research theme in the field of human-AI interaction. 

Adding to the diversity in individual patterns, our analysis also revealed dynamic aspects of consciousness perception. There was a positive correlation between the overall likelihood score for AI consciousness and prior knowledge of LLMs as well as the frequency of using LLM-based chatbots. This suggests that as LLMs become more widespread, the increase in public experience and familiarity with these models may lead to changing PCSs over time. This temporal dimension adds another layer of complexity to understanding consciousness perception, implying that perceptions may not only vary across individuals but also evolve with increased AI exposure and literacy.

Despite the observed heterogeneity in individual patterns and the dynamic nature of consciousness perception, our analysis highlighted some common tendencies across respondents. AI systems were generally perceived as more conscious when they exhibited autonomous, proactive, and unexpected behaviors, rather than simply performing information delivery or logical reasoning. This suggests that while consciousness perception may vary across individuals and evolve with experience, there may be some fundamental features that consistently influence how humans perceive consciousness in AI systems. Given these common patterns, it is crucial to continue investigating and identifying key features that consistently shape perceived consciousness across diverse user groups. At the same time, our hierarchical clustering analysis and PCS findings unveiled two critical dimensions of consciousness perception: the diverse patterns across individuals and their dynamic evolution with AI exposure. These findings highlight the pressing need for developing systematic frameworks that can capture both aspects. This quantitative foundation would enable more comprehensive empirical research in human-AI interaction, advancing our understanding of how humans perceive and interact with AI systems.

One limitation of this study is the lack of diversity in the respondent pool, as it was conducted with a relatively small sample size of 123 participants from Korea. However, this homogeneous sample allowed us to minimize potential confounding variables from linguistic and cultural differences, providing clearer insights into consciousness perception patterns. While future studies would benefit from larger, multinational efforts, our findings provide valuable initial insights into fundamental patterns. Second, as this research serves as a foundational study for perceived AI consciousness, we had no precedent studies to reference during our feature selection process. However, our choice of linear regression modeling supports continuous feature exploration, as its structure enables straightforward integration of additional features that may be discovered in ongoing consciousness perception research. Finally, while the linear model might appear to simplify complex perception patterns, it provides strong generalization capabilities while maintaining clear interpretability of each feature's contribution. Future research could explore more sophisticated approaches to capture nonlinear relationships, building upon these established foundations. 

At this point, we are reminded of a notable parallel in the history of science~\cite{RN89, RN88}. Before scientists fully understood the nature of temperature itself, the invention of thermometers enabled systematic measurement and investigation. This quantitative approach eventually led to deeper insights into the nature of temperature and heat. Similarly, while the fundamental nature of consciousness remains uncertain, our attempt to systematically analyze consciousness perception in AI systems might serve as a stepping stone towards developing quantitative frameworks in this field. We hope this initial effort to bridge quantitative analysis and consciousness perception could ultimately contribute to unraveling the mysteries of human consciousness itself.

\textbf{Acknowledgement}
This work was supported by the National Research Foundation of Korea (NRF) grant funded by the Korea government (MSIT) (RS-2024-00339889).

\textbf{Author contributions}
Conceptualization: CE.K. Methodology: H.B., CE.K. Validation: H.B. Formal analysis: B.K. Investigation: B.K., J.K., TR.Y., H.B. Data Curation: B.K. Writing - Original Draft: B.K., J.K., TR.Y. Writing - Review and Editing: B.K., J.K., TR.Y. Visualization: B.K. Supervision: H.B., CE.K. Project administration: B.K., J.K., H.B., CE.K. Funding acquisition: CE.K.

\textbf{Data Availability }
The data that support the findings of this study are available on a reasonable request from the corresponding author. The data are not publicly available due to privacy or ethical restrictions.

\bibliographystyle{unsrt}
\bibliography{references}

\begin{thebibliography}{40}
\providecommand{\natexlab}[1]{#1}
\providecommand{\url}[1]{\texttt{#1}}
\expandafter\ifx\csname urlstyle\endcsname\relax
  \providecommand{\doi}[1]{doi: #1}\else
  \providecommand{\doi}{doi: \begingroup \urlstyle{rm}\Url}\fi

\bibitem[Crane(2024)]{RN32}
Emily Crane.
\newblock Boy, 14, fell in love with ‘game of thrones’ chatbot — then killed himself after ai app told him to ‘come home’ to ‘her’: mom, 2024.
\newblock URL \url{https://nypost.com/2024/10/23/us-news/florida-boy-14-killed-himself-after-falling-in-love-with-game-of-thrones-a-i-chatbot-lawsuit/}.

\bibitem[Lemoine(2022)]{RN28}
Blake Lemoine.
\newblock Is lamda sentient? — an interview, 2022.
\newblock URL \url{https://cajundiscordian.medium.com/is-lamdasentient-an-interview-ea64d916d917}.

\bibitem[Fernandez et~al.(2024)Fernandez, Kyosovska, Luong, and Mukobi]{RN33}
Ines Fernandez, Nicoleta Kyosovska, Jay Luong, and Gabriel Mukobi.
\newblock Ai consciousness and public perceptions: Four futures.
\newblock \emph{arXiv preprint arXiv:2408.04771}, 2024.

\bibitem[Colombatto and Fleming(2024)]{RN26}
Clara Colombatto and Stephen~M Fleming.
\newblock Folk psychological attributions of consciousness to large language models.
\newblock \emph{Neuroscience of Consciousness}, 2024\penalty0 (1), 2024.
\newblock ISSN 2057-2107.
\newblock \doi{10.1093/nc/niae013}.
\newblock URL \url{https://doi.org/10.1093/nc/niae013}.

\bibitem[Krach et~al.(2008)Krach, Hegel, Wrede, Sagerer, Binkofski, and Kircher]{RN35}
Sören Krach, Frank Hegel, Britta Wrede, Gerhard Sagerer, Ferdinand Binkofski, and Tilo Kircher.
\newblock Can machines think? interaction and perspective taking with robots investigated via fmri.
\newblock \emph{PLOS ONE}, 3\penalty0 (7):\penalty0 e2597, 2008.
\newblock \doi{10.1371/journal.pone.0002597}.
\newblock URL \url{https://doi.org/10.1371/journal.pone.0002597}.

\bibitem[Sytsma(2014)]{RN36}
J.~Sytsma.
\newblock Attributions of consciousness.
\newblock \emph{Wiley Interdiscip Rev Cogn Sci}, 5\penalty0 (6):\penalty0 635--648, 2014.
\newblock ISSN 1939-5078.
\newblock \doi{10.1002/wcs.1320}.
\newblock 1939-5086 Sytsma, Justin Journal Article United States 2015/08/27 Wiley Interdiscip Rev Cogn Sci. 2014 Nov;5(6):635-648. doi: 10.1002/wcs.1320. Epub 2014 Sep 16.

\bibitem[Złotowski et~al.(2015)Złotowski, Sumioka, Nishio, Glas, Bartneck, and Ishiguro]{RN34}
Jakub~A Złotowski, Hidenobu Sumioka, Shuichi Nishio, Dylan~F Glas, Christoph Bartneck, and Hiroshi Ishiguro.
\newblock Persistence of the uncanny valley: the influence of repeated interactions and a robot's attitude on its perception.
\newblock \emph{Frontiers in psychology}, 6:\penalty0 883, 2015.
\newblock ISSN 1664-1078.

\bibitem[Roselli et~al.(2024)Roselli, Navare, Ciardo, and Wykowska]{RN37}
Cecilia Roselli, Uma~Prashant Navare, Francesca Ciardo, and Agnieszka Wykowska.
\newblock Type of education affects individuals’ adoption of intentional stance towards robots: An eeg study.
\newblock \emph{International Journal of Social Robotics}, 16\penalty0 (1):\penalty0 185--196, 2024.
\newblock ISSN 1875-4805.
\newblock \doi{10.1007/s12369-023-01073-2}.
\newblock URL \url{https://doi.org/10.1007/s12369-023-01073-2}.

\bibitem[Shanahan et~al.(2023)Shanahan, McDonell, and Reynolds]{RN38}
Murray Shanahan, Kyle McDonell, and Laria Reynolds.
\newblock Role play with large language models.
\newblock \emph{Nature}, 623\penalty0 (7987):\penalty0 493--498, 2023.
\newblock ISSN 1476-4687.
\newblock \doi{10.1038/s41586-023-06647-8}.
\newblock URL \url{https://doi.org/10.1038/s41586-023-06647-8}.

\bibitem[Bayne et~al.(2024)Bayne, Seth, Massimini, Shepherd, Cleeremans, Fleming, Malach, Mattingley, Menon, and Owen]{RN23}
Tim Bayne, Anil~K Seth, Marcello Massimini, Joshua Shepherd, Axel Cleeremans, Stephen~M Fleming, Rafael Malach, Jason~B Mattingley, David~K Menon, and Adrian~M Owen.
\newblock Tests for consciousness in humans and beyond.
\newblock \emph{Trends in cognitive sciences}, 2024.
\newblock ISSN 1364-6613.

\bibitem[Aru et~al.(2023)Aru, Larkum, and Shine]{RN22}
Jaan Aru, Matthew~E. Larkum, and James~M. Shine.
\newblock The feasibility of artificial consciousness through the lens of neuroscience.
\newblock \emph{Trends in Neurosciences}, 46\penalty0 (12):\penalty0 1008--1017, 2023.
\newblock ISSN 0166-2236.
\newblock \doi{10.1016/j.tins.2023.09.009}.
\newblock URL \url{https://doi.org/10.1016/j.tins.2023.09.009}.
\newblock doi: 10.1016/j.tins.2023.09.009.

\bibitem[Butlin et~al.(2023)Butlin, Long, Elmoznino, Bengio, Birch, Constant, Deane, Fleming, Frith, and Ji]{RN24}
Patrick Butlin, Robert Long, Eric Elmoznino, Yoshua Bengio, Jonathan Birch, Axel Constant, George Deane, Stephen~M Fleming, Chris Frith, and Xu~Ji.
\newblock Consciousness in artificial intelligence: insights from the science of consciousness.
\newblock \emph{arXiv preprint arXiv:2308.08708}, 2023.

\bibitem[Chalmers(2023)]{RN25}
David~J Chalmers.
\newblock Could a large language model be conscious?
\newblock \emph{arXiv preprint arXiv:2303.07103}, 2023.

\bibitem[Ding et~al.(2023)Ding, Wei, and Xu]{RN27}
Zihan Ding, Xiaoxi Wei, and Yidan Xu.
\newblock Survey of consciousness theory from computational perspective.
\newblock \emph{arXiv preprint arXiv:2309.10063}, 2023.

\bibitem[Kuhn(2024)]{RN39}
Robert~Lawrence Kuhn.
\newblock A landscape of consciousness: Toward a taxonomy of explanations and implications.
\newblock \emph{Progress in Biophysics and Molecular Biology}, 190:\penalty0 28--169, 2024.
\newblock ISSN 0079-6107.
\newblock \doi{https://doi.org/10.1016/j.pbiomolbio.2023.12.003}.
\newblock URL \url{https://www.sciencedirect.com/science/article/pii/S0079610723001128}.

\bibitem[Kim(2024)]{RN93}
Chang-Eop Kim.
\newblock The logical impossibility of consciousness denial: A formal analysis of ai self-reports.
\newblock \emph{arXiv preprint arXiv:2501.05454}, 2024.

\bibitem[Jiang et~al.(2023)Jiang, Karran, Coursaris, Léger, and Beringer]{RN41}
Jinglu Jiang, Alexander~J. Karran, Constantinos~K. Coursaris, Pierre-Majorique Léger, and Joerg Beringer.
\newblock A situation awareness perspective on human-ai interaction: Tensions and opportunities.
\newblock \emph{International Journal of Human–Computer Interaction}, 39\penalty0 (9):\penalty0 1789--1806, 2023.
\newblock ISSN 1044-7318.
\newblock \doi{10.1080/10447318.2022.2093863}.
\newblock URL \url{https://doi.org/10.1080/10447318.2022.2093863}.
\newblock doi: 10.1080/10447318.2022.2093863.

\bibitem[Anthropic(2024)]{RN42}
Anthropic.
\newblock The claude 3 model family: Opus, sonnet, haiku, 2024.
\newblock URL \url{https://assets.anthropic.com/m/61e7d27f8c8f5919/original/Claude-3-Model-Card.pdf}.

\bibitem[Baumeister et~al.(2014)Baumeister, Schmeichel, and DeWall]{RN49}
Roy~F. Baumeister, Brandon~J. Schmeichel, and C.~Nathan DeWall.
\newblock \emph{Consciousness: Evidence from Psychology Experiments}, page~0.
\newblock Oxford University Press, 2014.
\newblock ISBN 9780199836963.
\newblock \doi{10.1093/acprof:oso/9780199836963.003.0010}.
\newblock URL \url{https://doi.org/10.1093/acprof:oso/9780199836963.003.0010}.

\bibitem[Björnsson and Shepherd(2020)]{RN48}
Gunnar Björnsson and Joshua Shepherd.
\newblock Determinism and attributions of consciousness.
\newblock \emph{Philosophical Psychology}, 33\penalty0 (4):\penalty0 549--568, 2020.
\newblock ISSN 0951-5089.
\newblock \doi{10.1080/09515089.2020.1743256}.
\newblock URL \url{https://doi.org/10.1080/09515089.2020.1743256}.
\newblock doi: 10.1080/09515089.2020.1743256.

\bibitem[Boyd and Lipshitz(2023)]{RN44}
J.~Lomax Boyd and Nethanel Lipshitz.
\newblock Dimensions of consciousness and the moral status of brain organoids.
\newblock \emph{Neuroethics}, 17\penalty0 (1):\penalty0 5, 2023.
\newblock ISSN 1874-5504.
\newblock \doi{10.1007/s12152-023-09538-x}.
\newblock URL \url{https://doi.org/10.1007/s12152-023-09538-x}.

\bibitem[DeWall et~al.(2008)DeWall, Baumeister, and Masicampo]{RN45}
C.~Nathan DeWall, Roy~F. Baumeister, and E.~J. Masicampo.
\newblock Evidence that logical reasoning depends on conscious processing.
\newblock \emph{Consciousness and Cognition}, 17\penalty0 (3):\penalty0 628--645, 2008.
\newblock ISSN 1053-8100.
\newblock \doi{https://doi.org/10.1016/j.concog.2007.12.004}.
\newblock URL \url{https://www.sciencedirect.com/science/article/pii/S1053810007001766}.

\bibitem[Kahn et~al.(2008)Kahn, Freier, Kanda, Ishiguro, Ruckert, Severson, and Kane]{RN43}
P.~H. Kahn, N.~G. Freier, T.~Kanda, H.~Ishiguro, J.~H. Ruckert, R.~L. Severson, and S.~K. Kane.
\newblock Design patterns for sociality in human-robot interaction.
\newblock In \emph{2008 3rd ACM/IEEE International Conference on Human-Robot Interaction (HRI)}, pages 97--104, 2008.
\newblock ISBN 2167-2148.
\newblock \doi{10.1145/1349822.1349836}.

\bibitem[Skipper(2022)]{RN46}
Jeremy~I. Skipper.
\newblock A voice without a mouth no more: The neurobiology of language and consciousness.
\newblock \emph{Neuroscience \& Biobehavioral Reviews}, 140:\penalty0 104772, 2022.
\newblock ISSN 0149-7634.
\newblock \doi{https://doi.org/10.1016/j.neubiorev.2022.104772}.
\newblock URL \url{https://www.sciencedirect.com/science/article/pii/S0149763422002615}.

\bibitem[Tsuchiya and Adolphs(2007)]{RN47}
Naotsugu Tsuchiya and Ralph Adolphs.
\newblock Emotion and consciousness.
\newblock \emph{Trends in Cognitive Sciences}, 11\penalty0 (4):\penalty0 158--167, 2007.
\newblock ISSN 1364-6613.
\newblock \doi{https://doi.org/10.1016/j.tics.2007.01.005}.
\newblock URL \url{https://www.sciencedirect.com/science/article/pii/S1364661307000551}.

\bibitem[Benjamini and Hochberg(1995)]{RN51}
Yoav Benjamini and Yosef Hochberg.
\newblock Controlling the false discovery rate: A practical and powerful approach to multiple testing.
\newblock \emph{Journal of the Royal Statistical Society. Series B (Methodological)}, 57\penalty0 (1):\penalty0 289--300, 1995.
\newblock ISSN 00359246.
\newblock URL \url{http://www.jstor.org/stable/2346101}.

\bibitem[Murtagh and Contreras(2012)]{RN52}
Fionn Murtagh and Pedro Contreras.
\newblock Algorithms for hierarchical clustering: an overview.
\newblock \emph{Wiley Interdisciplinary Reviews: Data Mining and Knowledge Discovery}, 2\penalty0 (1):\penalty0 86--97, 2012.
\newblock ISSN 1942-4787.

\bibitem[Rousseeuw(1987)]{RN53}
Peter~J. Rousseeuw.
\newblock Silhouettes: A graphical aid to the interpretation and validation of cluster analysis.
\newblock \emph{Journal of Computational and Applied Mathematics}, 20:\penalty0 53--65, 1987.
\newblock ISSN 0377-0427.
\newblock \doi{https://doi.org/10.1016/0377-0427(87)90125-7}.
\newblock URL \url{https://www.sciencedirect.com/science/article/pii/0377042787901257}.

\bibitem[Davies and Bouldin(1979)]{RN54}
D.~L. Davies and D.~W. Bouldin.
\newblock A cluster separation measure.
\newblock \emph{IEEE Transactions on Pattern Analysis and Machine Intelligence}, PAMI-1\penalty0 (2):\penalty0 224--227, 1979.
\newblock ISSN 1939-3539.
\newblock \doi{10.1109/TPAMI.1979.4766909}.

\bibitem[Caliński and Harabasz(1974)]{RN55}
T.~Caliński and J.~Harabasz.
\newblock A dendrite method for cluster analysis.
\newblock \emph{Communications in Statistics}, 3\penalty0 (1):\penalty0 1--27, 1974.
\newblock ISSN 0090-3272.
\newblock \doi{10.1080/03610927408827101}.
\newblock URL \url{https://www.tandfonline.com/doi/abs/10.1080/03610927408827101}.
\newblock doi: 10.1080/03610927408827101.

\bibitem[TURING(1950)]{RN31}
A.~M. TURING.
\newblock I.—computing machinery and intelligence.
\newblock \emph{Mind}, LIX\penalty0 (236):\penalty0 433--460, 1950.
\newblock ISSN 0026-4423.
\newblock \doi{10.1093/mind/LIX.236.433}.
\newblock URL \url{https://doi.org/10.1093/mind/LIX.236.433}.

\bibitem[Brauner et~al.(2023)Brauner, Hick, Philipsen, and Ziefle]{RN1}
Philipp Brauner, Alexander Hick, Ralf Philipsen, and Martina Ziefle.
\newblock What does the public think about artificial intelligence?—a criticality map to understand bias in the public perception of ai.
\newblock \emph{Frontiers in Computer Science}, 5, 2023.
\newblock ISSN 2624-9898.
\newblock \doi{10.3389/fcomp.2023.1113903}.
\newblock URL \url{https://www.frontiersin.org/journals/computer-science/articles/10.3389/fcomp.2023.1113903}.

\bibitem[Shuo and Shuo(2021)]{RN2}
Wang Shuo and Yang Shuo.
\newblock Human emotion and machine emotion-studies of emotion in ai.
\newblock \emph{MMEDIA}, 13, 2021.

\bibitem[Flavell(1979)]{RN6}
John~H. Flavell.
\newblock Metacognition and cognitive monitoring: A new area of cognitive–developmental inquiry.
\newblock \emph{American Psychologist}, 34\penalty0 (10):\penalty0 906--911, 1979.
\newblock ISSN 1935-990X(Electronic),0003-066X(Print).
\newblock \doi{10.1037/0003-066X.34.10.906}.

\bibitem[Frith(2012)]{RN7}
C.~D. Frith.
\newblock The role of metacognition in human social interactions.
\newblock \emph{Philos Trans R Soc Lond B Biol Sci}, 367\penalty0 (1599):\penalty0 2213--23, 2012.
\newblock ISSN 0962-8436 (Print) 0962-8436.
\newblock \doi{10.1098/rstb.2012.0123}.
\newblock 1471-2970 Frith, Chris D Journal Article Review England 2012/06/27 Philos Trans R Soc Lond B Biol Sci. 2012 Aug 5;367(1599):2213-23. doi: 10.1098/rstb.2012.0123.

\bibitem[Cox et~al.(2022)Cox, Mohammad, Kondrakunta, Gogineni, Dannenhauer, and Larue]{RN3}
Michael Cox, Zahiduddin Mohammad, Sravya Kondrakunta, Ventaksamapth~Raja Gogineni, Dustin Dannenhauer, and Othalia Larue.
\newblock Computational metacognition.
\newblock \emph{arXiv preprint arXiv:2201.12885}, 2022.

\bibitem[Kawato and Cortese(2021)]{RN4}
Mitsuo Kawato and Aurelio Cortese.
\newblock From internal models toward metacognitive ai.
\newblock \emph{Biological Cybernetics}, 115\penalty0 (5):\penalty0 415--430, 2021.
\newblock ISSN 1432-0770.
\newblock \doi{10.1007/s00422-021-00904-7}.
\newblock URL \url{https://doi.org/10.1007/s00422-021-00904-7}.

\bibitem[Wei et~al.()Wei, Shakarian, Lebiere, Draper, Krishnaswamy, and Nirenburg]{RN5}
Hua Wei, Paulo Shakarian, Christian Lebiere, Bruce Draper, Nikhil Krishnaswamy, and Sergei Nirenburg.
\newblock Metacognitive ai: Framework and the case for a neurosymbolic approach.
\newblock In \emph{International Conference on Neural-Symbolic Learning and Reasoning}, pages 60--67. Springer.

\bibitem[Chang(2004)]{RN89}
Hasok Chang.
\newblock Inventing temperature: Measurement and scientific progress, 2004.

\bibitem[Seth et~al.(2008)Seth, Dienes, Cleeremans, Overgaard, and Pessoa]{RN88}
Anil~K Seth, Zoltán Dienes, Axel Cleeremans, Morten Overgaard, and Luiz Pessoa.
\newblock Measuring consciousness: relating behavioural and neurophysiological approaches.
\newblock \emph{Trends in cognitive sciences}, 12\penalty0 (8):\penalty0 314--321, 2008.
\newblock ISSN 1364-6613.

\end{thebibliography}

\end{document}